\documentclass[superscriptaddress,twocolumn,amsmath,amssymb]{revtex4}

\usepackage{graphicx}
\usepackage{dcolumn}
\usepackage{bm}

\begin{document}

\preprint{}

\title{
Applicability of the Wong formula for fusion cross sections 
from light to heavy systems}
\author{N. W. Lwin} \affiliation{ Department
of Physics, Mandalay University, Mandalay, Myanmar}
\author{N. N. Htike} \affiliation{ Department
of Physics, Mandalay University, Mandalay, Myanmar}
\author{K. Hagino} \affiliation{ Department of Physics, Tohoku
  University, Sendai 980-8578, Japan}
\affiliation{ Research Center for Electron Photon Science,Tohoku University,
  1-2-1 Mikamine,  Sendai 982-0826, Japan}
\affiliation{National Astronomical Observatory of Japan, 2-21-1 Osawa, Mitaka,
   Tokyo 181-8588, Japan}


\begin{abstract}
We discuss the applicability of the Wong formula for fusion cross 
sections in a single-channel problem. 
To this end, we carry out a systematic study 
and compare the approximate fusion cross sections 
with the exact results in a wide mass region of reaction systems. 
We show that the deviation of the approximate results from the exact cross sections 
is large for light systems, even though the Wong formula 
provides a reasonable approximation for heavy systems. 
We also discuss the energy dependence of the deviation, and show that 
for a given projectile nucleus the critical energy, at which the deviation 
exceeds 5\% of the exact 
cross sections, increases as a function of the mass number of the 
target nucleus.  
\end{abstract}

\maketitle

\section{INTRODUCTION}
The nuclear fusion reaction 
is defined as a reaction to form a compound nucleus. 
It plays an essential role in several phenomena in physics, such as 
the energy production in stars, nucleosynthesis, and 
a synthesis of superheavy elements \cite{BT98,HT12,DHRS98,Back14}. 
Theoretically, the simplest approach to estimate fusion cross sections is to use 
the potential model, in which one assumes a spherical 
potential between the two colliding nuclei. In this model, 
the colliding nuclei are assumed to be spherical and inert, and 
fusion is 
simulated as an absorption of flux inside the Coulomb barrier. 
Fusion cross sections are then obtained by adding the penetrability of 
the Coulomb barrier for each partial wave. 

Approximating the Coulomb barrier with a parabolic function, Wong has 
derived a compact expression for fusion cross sections \cite{WONG}. 
The resultant formula, referred to as the Wong formula, has been 
widely used in the investigation of heavy-ion fusion reactions. 
For instance, the formula has been used to provide reference 
cross sections in order to discuss channel coupling effects. 
The formula has also been used recently in order to analyze 
fusion barrier distributions using a method based on the 
Bayesian spectral deconvolution \cite{H16}. 
Moreover, the correction to the Wong formula has been discussed in order 
to discuss the oscillatory behavior of fusion cross sections in 
symmetric and nearly symmetric light systems \cite{Poffe,RH15}. 

The original Wong formula tends to overestimate fusion cross sections in 
light systems. 
In order to cure this problem, the Wong formula has been generalized in 
Ref. \cite{RH15} by introducing an energy dependence to the barrier 
parameters. Even though the generalized Wong formula has been shown 
to reproduce well the exact results, it has yet to clarify 
from which system and from which energy 
the generalization becomes important. 
The aim of this paper is to perform a systematic study and address this question. 
To this end, we compare fusion cross sections obtained with the Wong formula 
with the exact results from light to heavy systems. We shall discuss the 
deviation from the exact cross sections 
as a function of the mass number of the target nucleus for a fixed projectile 
nucleus. 

The paper is organized as follows. 
In the next section, we will briefly 
present the formulation of the potential model for fusion reactions, 
and introduce the Wong formula. 
In Sec. III, we will apply the Wong formula to a multitude of systems 
and discuss its applicability. We will also introduce the critical 
energy, at which the deviation exceeds a certain fraction of the 
exact cross sections. We will finally summarize the paper in Sec. IV. 

\section{The Wong formula for fusion cross sections}

In the potential model, fusion cross sections 
are computed from a numerical solution of 
the radial Schr$\ddot{\textrm{o}}$dinger equation, 
\begin{eqnarray}
\left[-\frac{\hbar^2}{2\mu}\frac{d^2}{d r^2}+
      V_0(r)+\frac{l(l+1)\hbar^2}{2\mu {r}^2}-E\right]u_l(r)=0,
\label{Schrodinger} 
\end{eqnarray}
where $\mu$ is the reduced mass, $V_0(r)$ is an inter-nucleus potential 
(that is, a sum of a nuclear and the Coulomb potentials), 
$l$ is the relative angular momentum between the colliding nuclei, 
and $E$ is the incident energy in the center of mass frame. 
With the incoming wave boundary condition, this equation is solved 
with boundary conditions of \cite{HT12,ccfull} 
\begin{eqnarray}
u_l(r)&\sim& \sqrt{\frac{k}{k_l(r)}}\,T_l\,
\exp\left(-i\int^r_{r_{\rm min}} k_l(r')dr'\right), ~~~~~~r \leq r_{\rm min} 
\nonumber \\
\\
&=&
H_l^{(-)}(kr)-S_l\,H_l^{(+)}(kr),~~~~~r \to\infty.
\end{eqnarray}
Here, $H_l^{(\pm)}$ are the Coulomb wave functions, and $k$ and $k_l(r)$ 
are defined as $k=\sqrt{2\mu E/\hbar^2}$ and 
\begin{equation}
k_l(r)=\sqrt{\frac{2\mu}{\hbar^2}\left(E-V_0(r)
-\frac{l(l+1)\hbar^2}{2\mu r^2}\right)},
\end{equation}
respectively. $r_{\rm min}$ is the radius at which the incoming wave boundary 
condition is imposed and is taken somewhere 
inside the Coulomb barrier \cite{HT12,ccfull}. 
From the $S$-matrix, $S_l$, or the transmission coefficient, 
$T_l$, so obtained, fusion cross sections are evaluated as, 
\begin{eqnarray}
\sigma_{\rm fus}(E)=\frac{\pi}{k^2} \sum_{l=0}^{\infty}(2l+1)P_l(E),
\label{sigma-exact} 
\end{eqnarray}
with $P_l(E)=1-|S_l|^2=|T_l|^2$. 

In order to derive an analytic expression for fusion cross 
sections, Wong has first approximated the Coulomb barrier 
with an inverted parabola, that is, 
\begin{eqnarray}
V_{0}(r) \sim V_B-\frac{1}{2}\mu\Omega^2 (r-R_B)^2.
\label{parabola} 
\end{eqnarray}
In the original Wong formula, the barrier position, $R_B$, and 
the barrier curvature,
$\hbar \Omega$, are assumed to be independent of $l$ and are evaluated 
for the $s$-wave. 
The effective potential
for the $l$-th partial wave then reads, 
\begin{eqnarray}
  V_{0}(r)+\frac{l(l+1)\hbar^2}{2\mu r^2} \sim && V_B
  + \frac{l(l+1)\hbar^2}{2\mu {R_B}^2}\nonumber \\
  &&-\frac{1}{2}\mu\Omega^2 (r-R_B)^2.
\label{effectivepot} 
\end{eqnarray}
Using the Hill-Wheeler formula \cite{HILL-WHEELER}, 
the penetration probability, $P_l(E)$, is calculated as, 
\begin{eqnarray}
P_{l}(E)=\frac{1}{1+\exp\left[
\frac{2\pi}{\hbar\Omega}\left(V_B+\frac{l(l+1)\hbar^2}{2\mu R_{B}^2}-E\right)\right]}.
\label{penetrability} 
\end{eqnarray}
Replacing the summation 
in Eq.~(\ref{sigma-exact}) by the integral, that is, 
\begin{eqnarray}
  \frac{\pi}{k^2}\sum_{l=0}^{\infty} (2l+1)P_l(E)\rightarrow
  \frac{\pi}{k^2}\int_{0}^{\infty} dl \, (2l+1)P_{l}(E),
\label{sum-to-int} 
\end{eqnarray}
one finally obtains the well-known Wong formula given by 
\begin{eqnarray}
\sigma_{\textrm{fus}}(E)=\frac{\hbar\Omega}{2E}R_{B}^2 \,
\ln\left[ 1+\exp\left(\frac{2\pi}{\hbar\Omega} (E-V_{B})\right)\right].
\label{Wong} 
\end{eqnarray}
Notice that at energies well above the Coulomb barrier, i.e., 
$E-V_B\gg \hbar\Omega/2\pi$,
the Wong formula leads to the classical fusion cross section, 
\begin{eqnarray}
\sigma_{\textrm{fus}}(E)\sim \pi R_{B}^2 \left(1-\frac{V_{B}}{E}\right).
\label{classical} 
\end{eqnarray}
See Appendix B in Ref. \cite{HT12} for the performance of the 
parabolic approximation and the Wong formula for the $^{16}$O+$^{144}$Sm 
system. 

\begin{figure}[tb]
 \begin {tabular}{cc}
   \begin{minipage}{.2\textwidth}    
\scalebox{0.42}{\includegraphics[clip]{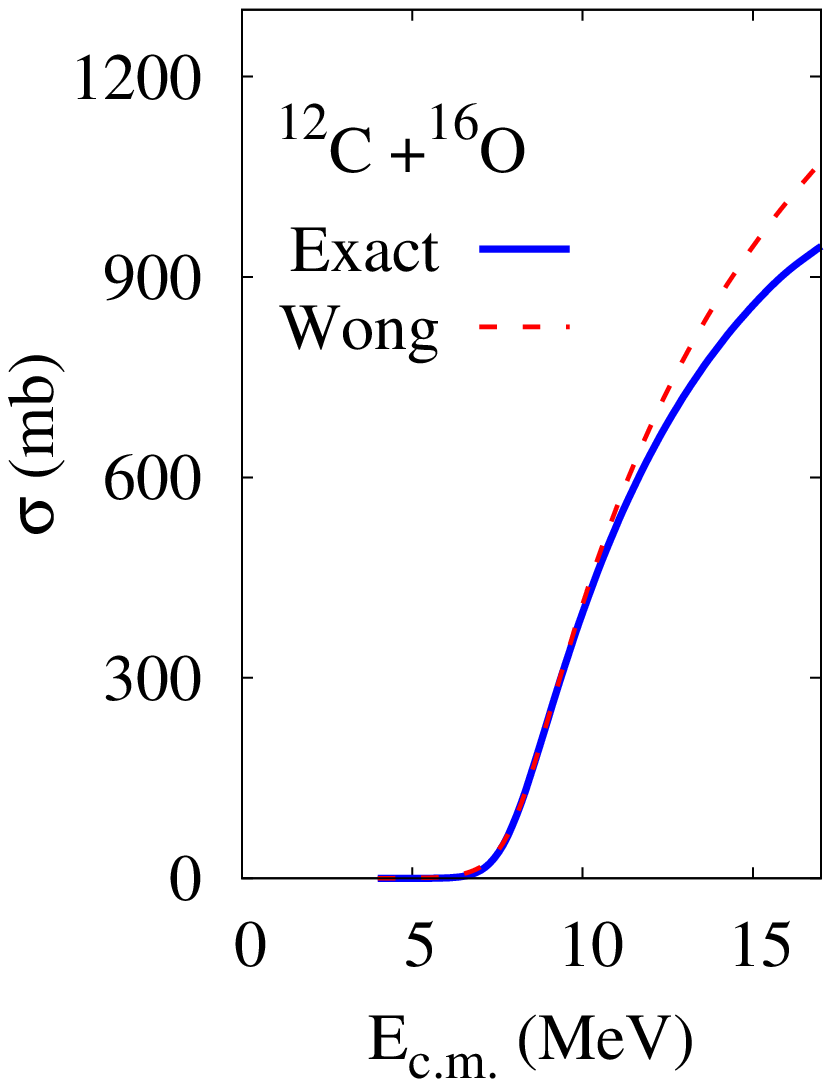}}  
\end{minipage}
\begin{minipage}{.2\textwidth}
\scalebox{0.42}{\includegraphics[clip]{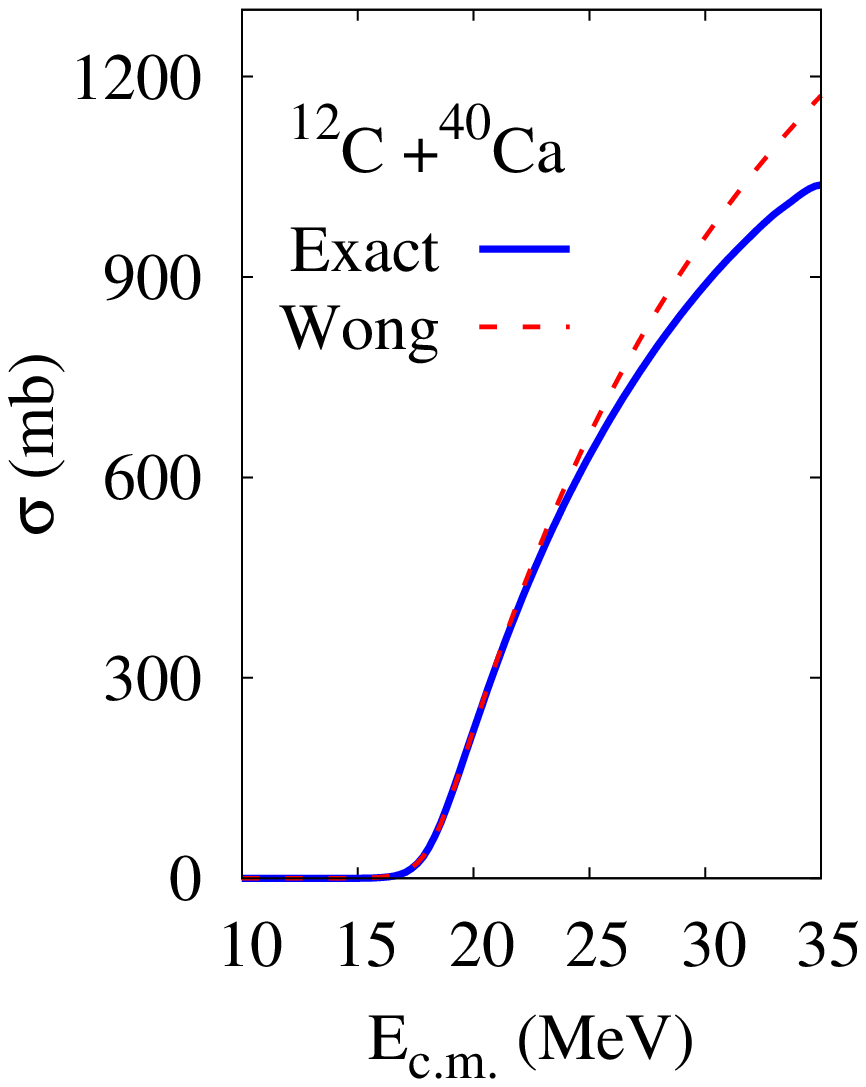}}
\end{minipage}
\end{tabular}
\end{figure}
\begin{figure}[htb]
\begin {tabular}{cc}
\begin{minipage}{.2\textwidth}
\scalebox{0.4}{\includegraphics[clip]{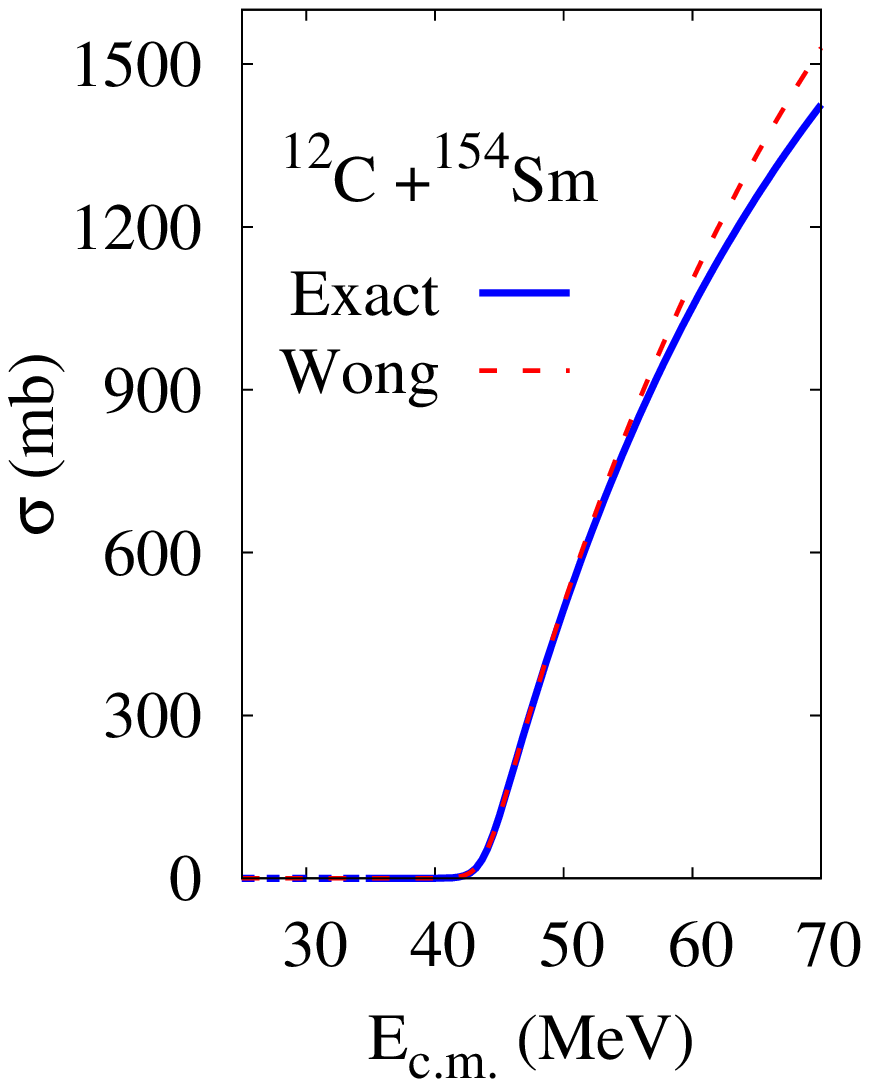}}
\end{minipage}
\begin{minipage}{.2\textwidth}
\scalebox{0.4}{\includegraphics[clip]{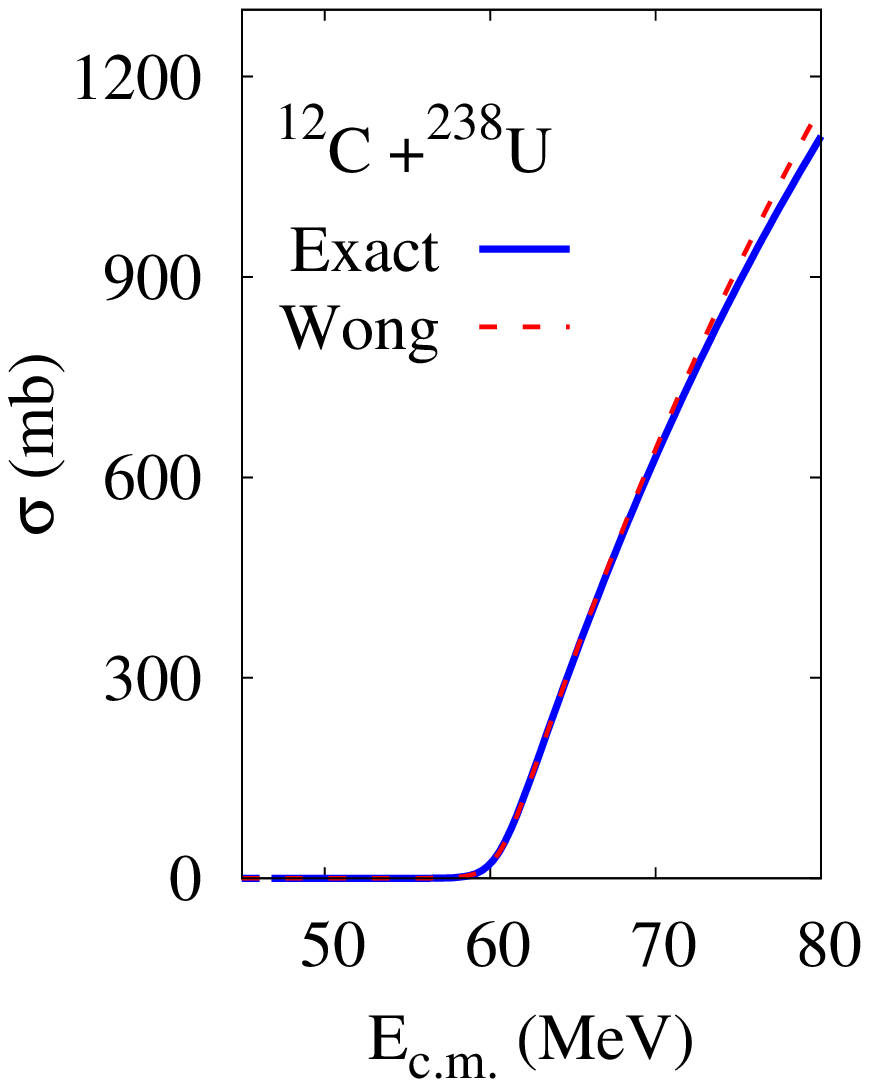}}
\end{minipage}
\end{tabular}
\caption{\label{light-to-heavy}
Comparisons between the exact (the solid line) and approximate (the dashed line) fusion cross sections 
for four selected systems. The approximate solutions are obtained with the 
Wong formula. The Aky\"uz-Winther potential is employed for the internuclear 
potential.}
\end{figure}

Figure~\ref{light-to-heavy} compares 
fusion cross
sections with 
the Wong formula (the dashed line) to 
those obtained by numerically solving the Schr\"odinger 
equation (the solid line), 
for
the $^{12}$C+$^{16}$O, $^{40}$Ca, $^{154}$Sm and $^{238}$U systems. 
To this end, 
we employ the 
Aky\"uz-Winther potential
\cite{AKYUZ} for a nuclear part of the nucleus-nucleus potential. 
As can be seen in the figure, 
the Wong formula tends to overestimate fusion cross sections, although 
the agreement with the exact results considerably improves as the mass 
number of the target increases \cite{RH15} (see also Ref. \cite{TCH17}). 

\section {Systematics}

In order to discuss more systematically the performance of the Wong formula, 
in this section we vary the target nucleus from C to U 
for a fixed projectile nucleus. 
For each target nuclide, we choose the isotope which has the
largest natural abundance. 
As the projectile nucleus, we choose $^4$He, 
$^6$Li, $^{12}$C, $^{16}$O, 
and $^{20}$Ne. 
In order to compare the deviation of the Wong formula in different 
systems, we introduce the following quantity: 
\begin{eqnarray}
\Delta \sigma \equiv \frac{\int^{E_{\rm max}}_{E_{\rm min}}
|\sigma_{\rm exact}(E)
-\sigma_{\rm Wong}(E)|dE}{\int^{E_{\rm max}}_{E_{\rm min}} \sigma_{\rm exact}(E)dE}. 
\label {delta-sigma}       
\end{eqnarray}
We choose the minimum and the maximum energies for the integration 
to be $E_{\rm min}$= 0.9 $V_B$ and
$E_{\rm max}$ = 1.1 $V_B$, respectively. 

\begin{figure}[tb]
\begin{center}\leavevmode
\includegraphics[width=0.91\linewidth, clip]{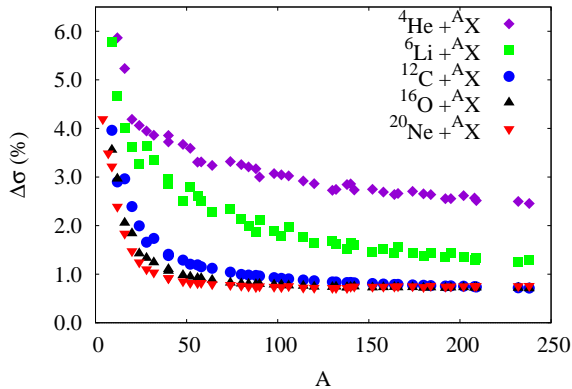}
\end{center}
\caption{\label{sigma-A}
The deviation from the exact results, defined by Eq. (\ref{delta-sigma}), 
of the fusion cross sections obtained with the 
Wong formula. 
This 
quantity is plotted 
as a function of the mass number of the target nucleus for five projectile 
nuclei indicated in the figure. }
\end{figure}

Figure~\ref{sigma-A} shows the calculated results as a function
of the mass number $A$ of the target nucleus. 
We again use the Aky\"uz-Winther potential for each system. 
One can see that the deviation, $\Delta\sigma$, 
decreases as a function of the mass number of the target nucleus. 
For the 
$^{12}$C, $^{16}$O, and $^{20}$Ne projectiles, 
the deviation 
almost saturates at around $A\sim 50$, that corresponds to 
$Z \sim$ 20. 
On the other hand, for the 
lighter projectiles, the decrease of $\Delta\sigma$ is much slower 
and especially for $^4$He the deviation is relatively large even for a heavy target. 
This implies that the original Wong formula is not applicable for 
light projectiles such as $^4$He, 
and the generalization proposed in Ref. \cite{RH15} is important. 
For heavier projectile nuclei, 
as we have noted, the original Wong formula works well 
as long as the mass number of the target nucleus 
is around $A\sim$ 50 or larger. 

\begin{figure}[tb]
\begin{center}\leavevmode
\includegraphics[width=0.91\linewidth, clip]{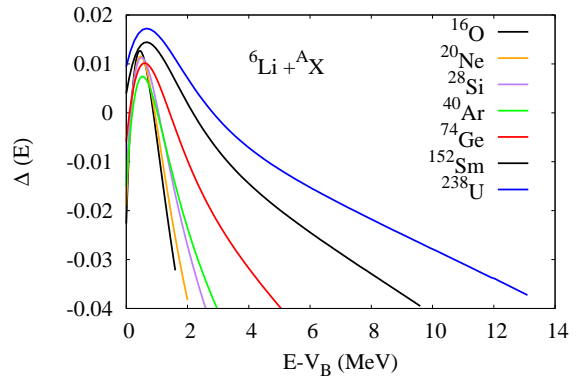}
\end{center}
\caption{\label{delta-E}
The deviation of the Wong formula given by Eq. (\ref{delta}) for 
$^6$Li induced fusion reactions as a function of the incident energy 
relative to the barrier height for each system.  
The target nucleus is $^{238}$U, $^{152}$Sm, $^{74}$Ge, $^{40}$Ar, $^{28}$Si, 
$^{20}$Ne, and $^{16}$O in the decreasing order at $E-V_B$ = 4 MeV. } 
\end{figure}

In order to see the energy dependence of the deviation of the Wong formula, 
we next define 
a quantity $\Delta(E)$ 
\begin{eqnarray}
  \Delta(E)  \equiv  \frac{\sigma_{\rm exact}(E)-\sigma_{\rm Wong}(E)}
          {\sigma_{\rm exact}(E)}, 
\label {delta} 
\end{eqnarray}
which indicates the degree
of deviation at a given energy, $E$. 
Fig.~\ref{delta-E} shows the quantity $\Delta(E)$ 
for the $^6$Li projectile on several target nuclei 
from O to U, as a function of energy relative to the barrier 
height for each system. 
Except for the energy region just above the barrier, 
$\Delta(E)$ is negative, that is, the Wong formula overestimates 
fusion cross sections. 
Furthermore, the slope
of $\Delta(E)$ is much steeper for the lighter targets 
compared to that for the heavier targets, that is consistent with the 
finding shown in Fig. \ref{sigma-A}. 

At energies slightly above the barrier,
the deviation $\Delta(E)$ is positive, thus the 
Wong formula underestimates fusion cross sections. 
This is due to 
the fact 
that the parabolic approximation underestimates the width 
of the Coulomb barrier because of the asymmetric shape of the Coulomb 
barrier caused by 
the long range Coulomb interaction (see Fig. 15 in Ref. \cite{HT12}). 
This leads to a smaller tunneling probability at energies above the 
barrier, thus reducing fusion cross sections.
At higher energies, on the other hand, the tunneling is not
important, and the $l$-dependence of the barrier position
$R_{\rm B}$ comes into a play to reduce fusion cross sections 
from the approximate cross sections \cite{RH15}. 

\begin{figure}[tb]
\begin{center}\leavevmode
\includegraphics[width=0.91\linewidth, clip]{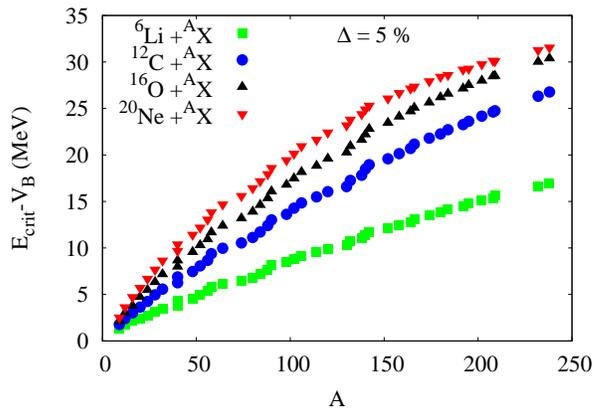}
\end{center}
\caption{\label{Ecrit-EB}
The critical energy, $E_{\rm crit}$, at which the deviation defined by Eq. (\ref{delta}) 
exceeds 5\%, measured relative to the barrier height for each system as a function of the 
mass number of the target nucleus. } 
\end{figure}

Let us next define the critical energy $E_{\rm crit}$ as the energy at which 
the absolute value of the deviation, $|\Delta(E)|$, exceeds a certain value. 
We arbitrarily choose it to be 5\%. 
Fig.~\ref{Ecrit-EB} shows the critical energy with respect to 
the barrier height for each system as a function of the mass number 
of the target nucleus. For the projectile, we choose 
$^6$Li, $^{12}$C, $^{16}$O, and $^{20}$Ne. 
The figure shows that for light systems 
the critical energy is reached already at an energy 
slightly above the barrier, while 
one has to go up to a higher energy for
heavy systems. For instance, the critical energy is 3.02 MeV higher than 
the barrier for the $^{12}$C+$^{16}$O system whereas it is 30.37 MeV higher 
than the barrier for the 
$^{12}$O+$^{238}$U system.
We notice that this is consistent with what we have shown in Fig. \ref{light-to-heavy}. 

\section{Summary}

We have carried out a comparative study on the applicability of the Wong 
formula for single-channel fusion cross sections. 
To this end, we have compared the fusion cross sections obtained with the 
Wong formula to the exact results from light to heavy systems. 
We have shown that the Wong formula leads to reasonable results 
when the target 
nucleus is in the $A\sim$ 50 region or heavier for the projectiles of 
$^6$Li, $^{12}$C, $^{16}$O, and $^{20}$Ne, while the deviation is still 
large 
even for a heavy target in systems with $^4$He projectile. 
We have also investigated the energy dependence of the deviation, and have 
shown that the deviation quickly becomes large for light systems 
as the energy increases while a high incident energy is required for heavy 
systems before the deviation becomes significant. 
These conclusions indicate that a care must be taken in using the 
Wong formula in light systems, for which the generalization discussed 
in Ref. \cite{RH15} becomes essential.

\end{document}